\documentclass[prl,aps,twocolumn,showpacs]{revtex4}
\usepackage{graphicx}

\newcommand{\half}{{\textstyle{1\over 2}}}
\newcommand{\be}{\begin{equation}}
\newcommand{\ee}{\end{equation}}
\newcommand{\beqa}{\begin{eqnarray}}
\newcommand{\eeqa}{\end{eqnarray}}
\newcommand{\nonu}{\nonumber}
\newcommand{\re}{{\rm e}}
\newcommand{\pr}{\prime}

\begin{document}
  
  \title{Equilibration of a dissipative quantum oscillator.}

   \author{Vinay Ambegaokar}
\affiliation{Laboratory of Atomic and Solid State Physics, Cornell 
University, Ithaca, New York 14853}
\date{May 25,  2006}
\pacs{03.75.Ss}

 \begin{abstract}
An explicit demonstration is given of a harmonic oscillator in equilibrium approaching the equilibrium of a corresponding interacting system by coupling it to a thermal bath consisting of a continuum of harmonic oscillators.
  \end{abstract}
   \maketitle

In the theory of interacting many-particle systems, an initial condition of thermal equilibrium for the system without interactions is often imposed as a convenient device, with the implicit assumption that time-evolution via the full Hamiltonian will lead to the correct correlation functions.  Such ``switching on" of interactions goes back a long time \cite{BK, LK}, and is also a part of more recent discussions  \cite{ DL, RS, AK}. In one form or another, the assumption is to be found in most theories of transport under non-equilibrium conditions \cite{SH}. However, and despite its ubiquity, it is rarely \cite{LC} addressed, and sometimes questioned \cite{NGK}.  The straightforward and complete demonstration within a simple model offered here may therefore be of general interest. 

To set the calculation in its general context, consider a system consisting of a single particle ``sub-system," a many-particle ``bath," and interactions between the two.  Let the Hamiltonian for the interacting system be $H$, and let $H_0$ be the Hamiltonian for the complete system  with the particle-bath interactions omitted.  Now consider two reduced density matrices, one describing true (time-independent) equilibrium, and a second corresponding to the time-evolution of the equilibrium of the uncoupled system and bath.  These may be written as
 \be\label{1d}
\rho(Q_{0}^\prime, Q_{0}^{\prime\prime}) = Tr\{\vert Q_{0}^{\prime\prime}\rangle\langle Q_{0}^\prime\vert ~\rho_{th}  \},~~~\rm{and}
\ee 
\be 
 \rho_0(Q_{0}^\prime, Q_{0}^{\prime\prime},t) = Tr\{\vert Q_{0}^{\prime\prime}\rangle\langle Q_{0}^\prime\vert~ \re ^{-i H t}  \rho^0_{th}  \re^{+i H t} \}.
\ee
In these equations, the primed quantities refer to eigenvalues of the sub-system position operator $Q_0$,   $Tr$ indicates a trace over all states of $H$,  $\rho_{th} $ is  $\exp\{-\beta H \} / Tr\exp\{-\beta H\}$ with $\beta$ the reciprocal temperature, and $\rho_{th}^0 $  is identically defined except that the $H$  of
 Eq.(1) is replaced by $H_0$.  The question being posed  is whether the quantity defined in (2) becomes (1) for large times.

The model that will be treated here is widely used to describe a dissipative quantum oscillator:  
 \be
 H =\half  [P_0^2 +\Omega_0^2 Q_0^2 ]+ \half \sum_k \{P_k^2 + \Omega_k^2[Q_k + (\alpha_k Q_0/\Omega_k^2)]^2\} ,
 \ee
 with $H_0\equiv H(\alpha_k \rightarrow 0).$ Here the oscillator $0$ is the subsystem, and the others the bath.
 
Because the couplings in Eq.(3) are quadratic in the co-ordinates $ Q_i $, where $i$ is $0$ or any one of the $k$s, the problem is  formally solved by an orthogonal normal-mode transformation \cite{RJR} to new canonical variables $q_\nu, p_\nu$,
 \be
 Q_i= \sum_\nu X_{i\nu} q_\nu ~~~~~~~~~~~q_\nu= \sum_i X_{i \nu}  Q_i ~,
 \ee 
 with identical connections between $P_i$ and $p_\nu$.
 The normal-mode frequencies, call  them $\omega_\nu$,  and the transformation matrices $X$ are obtainable \cite{ FLO, NGK, VA} from a Green function whose inverse is given by 
 \be 
 g^{-1}(z) = z^2 - \Omega_0^2 - \sum_k \alpha_k^2 \big ( {1\over z^2 - \Omega_k^2} + {1\over \Omega_k^2} \big).
  \ee 
   
By putting Eq.(5) on a common denominator, and noting that the numerator is then the determinant whose zeros are the normal mode frequencies, one sees that, as a function of the complex variable $z$, $g$ has the representation \cite{FLO}
\be
g(z) = {\prod_k (z^2 - \Omega_k ^2)\over \prod_{\nu} (z^2 - \omega_\nu ^2)}.
\ee 
In words,  it has poles on the real axis at the normal-mode frequencies, and (interleaved) zeros also on the real axis at the bath frequencies.  Furthermore, \cite{ FLO, NGK, VA}
 \beqa
 {1\over X_{0\nu}} &=& \sqrt{1 + \sum_k\alpha_k^2/(\Omega_k^2 - \omega_\nu^2)^2}\nonu\\
 &=&\Big(\sqrt{{1\over 2 z} {\partial g^{-1}\over\partial z}}\Big)_{z=\omega_\nu}~~~{\rm and}\nonu\\
X_{k \nu} &=& {\alpha_k\over \omega_\nu^2 - \Omega_k^2 } X_{0\nu}.
 \eeqa

It is convenient to express the projection operator in Eqs.(1) and (2) as
\beqa
\vert Q_{0}^{\prime\prime}\rangle\langle Q_{0}^\prime\vert&= &\int _{-\infty}^\infty d u d v f(u,v) \re^{iP_0 u} \re^ {iQ_0 v}\nonu~~~~~ {\rm with}\\ 
f(u, v) &= &{1\over 2\pi} \re^{-i Q_{0}^\prime v} \delta(Q_{0}^\pr - Q_{o}^{\pr\pr} -u),
\eeqa
proved by taking matrix elements of both sides, to see that  I am here asking whether and, if so, how
\be
\mathcal A_0(t)\equiv\langle \re^{iP_0(t)u}\re^{iQ_0(t) v}\rangle_0~\longrightarrow~\mathcal A\equiv\langle \re^{iP_0u}\re^{iQ_0 v}\rangle.
\ee
On the left, the operators are in the Heisenberg picture corresponding to $H$ and the average is with respect to $\rho_{th}^0$; on the right, the time-independent average  is in $\rho_{th}$.  This latter quantity is easily evaluated using the method based on the Debye-Waller identity introduced in ref.\cite{VA} to obtain
\be
\ln \mathcal A = -\half u^2 \langle P_0^2\rangle -\half v^2\langle Q_0^2\rangle+{\textstyle{i\over 2}} uv,
\ee 
For $\mathcal A_0(t)$, note that the time development due to the full $H$ for $P_0$, $Q_0$
is most directly calculated by  using Eq.(4) to transform to the normal-mode variables $p_\nu$, $q_\nu$, evolving these according to the quantum mechanical (and classical) equations of motion
\beqa
p_\nu(t)&=& p_\nu (0) \cos (\omega_\nu t) - q_\nu (0)\omega_\nu\sin(\omega_\nu t),\nonu\\
q_\nu(t)&=&q_\nu (0) \cos (\omega_\nu t) + p_\nu (0) {1\over\omega_\nu}\sin(\omega_\nu t),
\eeqa
and then transforming back, using the second part of Eq.(4).   The Debye-Waller identity may again be invoked, because the Schr\"odinger, i.e.\ $t=0$, $P_i$s and $Q_i$s are independent oscillator variables for $H_0$, to obtain
\begin{widetext}
\beqa
\ln\mathcal A_0(t) = &-&\half u^2 \langle\big [\sum_i (\mathcal X_i^c P_i -\bar{\mathcal X}_i^s Q_i)\big]^2\rangle_0-\half v^2\langle\big [\sum_i (\mathcal X_i^c Q_i +{\mathcal X}_i^s P_i)\big]^2\rangle_0\nonu\\
&-&uv\langle\big [\sum_i (\mathcal X_i^c P_i-\bar{\mathcal X}_i^s Q_i)\big] \big[\sum_i (\mathcal X_i^c Q_i +{\mathcal X}_i^s P_i)\big]\rangle_0.
\eeqa
 Here the $\mathcal X$s are the time-dependent functions
\be
\mathcal X_i^c(t) \equiv \sum_\nu X_{0 \nu} X_{i \nu} \cos( \omega_\nu t), ~~ \mathcal X_i^s(t)\equiv\sum_\nu X_{0 \nu} X_{i \nu} {1\over\omega_ \nu} \sin( \omega_\nu t),~~ \rm{and}~\bar\mathcal X_i^s(t)\equiv\sum_\nu X_{0 \nu} X_{i \nu}~ {\omega_\nu} \sin( \omega_\nu t).
\ee
The double sums in Eq.(12) collapse to single sums, again because the oscillators $i$ are independent within $H_0$, yielding
\beqa
\ln \mathcal A_0(t) =& - &\half u^2\sum_i\big[(\mathcal X_i^c)^2\langle P_i^2\rangle_0 +(\bar{\mathcal X}_i^s)^2 \langle Q_i^2\rangle_0\big] - \half v^2 \sum_i\big[(\mathcal X_i^c)^2\langle Q_i^2\rangle_0 +(\mathcal X_i^s)^2 \langle P_i^2\rangle_0\big]\nonu\\
&-&u v \sum_i \big[\mathcal X_i^c \mathcal X_i^s \langle P_i^2\rangle_0-\bar\mathcal X_i^s \mathcal X_i^c \langle Q_i^2\rangle_0 -\textstyle{{i\over 2}}(\mathcal X_i^c)^2 - \textstyle{{i\over 2}}\bar \mathcal X _i^s\mathcal X_i^s\big].
\eeqa
\end{widetext}
Thus, Eq.(9) is true if the time-dependent coefficients of $u^2,$ $v^2,$ and $uv$ in Eq.(14) approach the corresponding time-independent ones in Eq.(10).

The sums over $\nu$ in Eq.(13) are most easily done using the method of refs.\cite{FLO, NGK, VA}.   In particular,
\beqa
 &\sum_\nu& X_{0\nu}^2 F(\omega_\nu) =2\oint {d z\over 2\pi i} z g(z) F(z)\nonu\\  &=&2\Im \int_0^{\omega_c} {d \omega\over
\pi} \omega g(\omega - i0^+) F(\omega).
\eeqa
The contour surrounds the real axis, where the function F has been assumed to be regular and zero for $\omega < 0$.   In the last form the upper limit   $ \omega_c$ is the highest normal-mode frequency, and the integral has been written as the imaginary part of the forward portion of  the contour.  Further, using Eq.(7)
\beqa
 &\sum_\nu& X_{0\nu}X_{k\nu} F(\omega_\nu) = 2\alpha_k\oint {d z\over 2\pi i} z {g(z)\over (z^2 - \Omega_k^2)} F(z)\nonu\\  &=&2\alpha_k\Im \int_0^{\omega_c} {d \omega\over
\pi} \omega {g(\omega - i0^+)\over (\omega -i0^+)^2 - \Omega_k^2} F(\omega).
\eeqa
Note that there is no contribution from the poles at $z= \pm \Omega_k$ because of the zeros in $g(z)$ at these points, evident from Eq.(6).  [For an amusing illustration of the importance of these zeros, see the final paragraph.]

Further progress requires a specification of the spectrum of bath oscillators.  On general grounds, one would expect that to damp out oscillations a continuous spectrum is needed \cite{DET}.  When this is the case it is meaningful to write
\be
 {\pi\over 2} \sum_k (\alpha_k^2/ \Omega_k)\delta (\omega -\Omega_k) = J(\omega),
 \ee
 where  $J(\omega)$ is a smooth function of its argument in the region $0 < \omega <\bar\omega_c $ ($\bar\omega_c$ is the highest bath frequency,) and zero elsewhere.   Substituting this form into Eq.(5), and converting the k sum into an integral, yields
 \be
 g^{-1} (\omega - i0^+)= \omega^2 - \Omega_0^2 -2 \int_0^{\bar\omega_c} {d\bar\omega\over \pi}{J(\bar\omega)\over \bar\omega}
 {\omega^2\over (\omega - i0^+)^2 - \bar\omega ^2}, 
\ee 
 implying $\Im g^{-1} (\omega - i0^+) = - J(\omega).$
 
 The time independent part of Eq.(14) may now be extracted.  [Here I will not examine the remaining time-dependent terms, leaving that to further analysis, below, of a more specific model.]  From Eqs.(13) and (16),
 \beqa
 \mathcal X_k^c(t) &=& 2\alpha_k\Im \int_0^{\omega_c} {d \omega\over
\pi} \omega {g(\omega - i0^+)\over (\omega -i0^+)^2 - \Omega_k^2} \cos(\omega t)\nonu\\
&=&\alpha_k\Re g(\Omega_k-i0^+) \cos(\Omega_k t) \nonu\\&~&+\alpha_k P \int_{-\omega_c}^{\omega_c} {d \omega\over
\pi} \omega {\Im g(\omega - i0^+)\over \omega^2 - \Omega_k^2} \Re e^{i\omega t}.
\eeqa
In the last line a factor of 2 has been removed and the evenness of the integrand used to change the limits of  the principal value integral (denoted by P.)  There is enough convergence to set $\omega_c\rightarrow \infty$.  Now, perform the principal value integral by adding one-half of the integral just below to one half  of the integral just above the real $\omega$
 axis. [Note that the scale of $\Im g(\omega - i0^+)$ is set by $J$, which is not small.]  Close the integral in the upper half plane to see that
\beqa
 \mathcal X_k^c(t) = \alpha_k \big[ \Re g(\Omega_k-i0^+) \cos(\Omega_k t)&&\nonu\\ - \Im g(\Omega_k -i0^+)\sin(\Omega_k t)\big]+ &&\dots,
 \eeqa
where the $\dots$ indicate the time-dependent contributions of the singularities of $\Im g(\omega - i0^+)$.   The distance of the nearest of these singularities from the real axis sets the longest time-scale of decaying terms.   In a similar way one finds
\beqa
\bar\mathcal X_k^s(t)= \alpha_k  \Omega_k\big[ \Re g(\Omega_k-i0^+) \sin(\Omega_k t)&&\nonu\\ +\Im g(\Omega_k-i0^+)\cos(\Omega_k t)\big] +&&\dots .
\eeqa

Consider the coefficient of $u^2$ in Eq.(14).  The term with $i=0$ is not relevant for the present purposes.  (See below.) Use the thermal averages
\be
\langle P_k^2 \rangle_0 = \Omega_k^2 \langle Q_k^2 \rangle_0 = {\Omega_k\over 2} \coth {\beta\Omega_k\over 2},
\ee
to see that the contribution of Eqs.(20,21) for all $k$ to the first term on the right hand side of Eq.(14) is the time-independent sum
\be
-{\half u^2} \sum_k \alpha_k^2 {\Omega_k\over 2} \coth{\beta\Omega_k\over 2} \big\vert g(\Omega_k - i0^+)\big \vert^2.
\ee

As the final step, use Eq.(17) to convert the sum over k to an integral and exploit $ -|z|^2 \Im z^{-1}= \Im z$ for any complex number to re-express  Eq.(23) as
\be
{-\half} u^2  \int_0^{\bar\omega_c} {d\omega\over \pi} 2\omega~ \Im g(\omega - i0^+) {\omega\over 2} \coth{\beta\omega\over 2}.
\ee
But this is exactly  what Eq.(15) gives for the first term on the right hand side of Eq.(10), provided that the band width of  the normal modes, $\omega_c$,  is the same as that of the bath, $\bar\omega_c$.  Of course, if this condition were not satisfied, the system would not be embedded in the bath and could not be expected to equilibrate with it.  

In a similar way, one can show that the other terms in Eq.(14) contain time-independent pieces which reproduce those in Eq.(10).  

Although there is no time dependence whatsoever in these pieces extracted from Eq.(14), it is of course not the case that the remainder does not contribute significantly at $t=0$.  In this limit, because of the sinusoidal functions, only $\mathcal X_i^c$ survives.  The completeness condition obeyed by the orthogonal transformation then requires
\be
\mathcal X_i^c (t=0) = \sum_v X_{0\nu} X_{i\nu} = \delta_{i 0},
\ee
making $\mathcal A_0$ look identical to $\mathcal A$, with the important difference that the average is in the uncoupled system.

As  mentioned below Eq.(20), the time-decaying parts of $\mathcal A_0(t)$ come from the singularities of $\Im g(\omega - i0^+)$, and thus depend on the specifics of the bath.  It may therefore be most instructive to treat a simple model.   Consider  ``ohmic" or linear dissipation \cite{LC}, in which $J(\omega)= \eta\omega$ for $0 < \omega < \bar\omega_c$.  In this situation Eqs.(17) and (5) give
\beqa
g^{-1}(z)&=& z^2 - \Omega_0^2 - ({\eta z\over\pi})\ln\big({z+\bar\omega_c\over z - \bar\omega_c}\big)\nonu\\ \rightarrow&~& z^2 - \Omega_0^2 + i\eta z~ \rm {sgn} (\Im z),~~ |z| \ll \bar\omega_c,
\eeqa
so that
\be
\Im g(\omega - i0^+) = {\eta \omega\over( \omega^2 -\Omega_0^2)^2 + \omega^2 \eta ^2}.
\ee
This function has simple poles at the four points $\pm \Omega \pm i\eta/2$, where $\Omega^2 = \Omega_0^2- \eta^2/4$.  It is  these poles that contribute to the time dependence of the $i=0$ terms in Eq.(13) and to the parts indicated by the $\dots$s in Eqs.(20) and (21).  For an underdamped oscillator, all of them thus decay like $\exp(-\eta t/2)$, showing in this case that the damping constant $\eta$ determines the rate at which equilibrium is reached.   

This completes the demonstration promised in the abstract to this paper.

As an illustration of the importance of the zeros of $g(z)$ in the evaluation of sums over normal mode frequencies, consider the following proof of the completeness relation $\sum_\nu X_{k\nu}^2 = 1.$
Using the method outlined in Eqs.(15) and (16) one obtains
\be
\sum_\nu X_{k\nu}^2 =2 \alpha_k^2 \oint {dz\over 2\pi i} z {g(z)\over (z^2 - \Omega_k^2)^2 }- {\alpha_k^2\over 2 \Omega_k} \big({\partial g\over\partial z}\big )_{z=\Omega_k}.
\ee
The subtracted term is the residue of the simple pole at $z=\Omega_k$ which remains after taking into account the zero at the same point in $g(z)$.  The contour integral can be seen to be zero.  The second term can be evaluated by using an identity expressing the invariance of the determinant of the dynamical matrix for the coupled oscillator system, referred to below Eq.(5), under the orthogonal normal mode transformation,
\be 
(z^2 - \bar\Omega_0^2)\prod_{\bar k } (z^2 - \Omega_{\bar k}^2) - \sum_{\bar k}\alpha_{\bar k}^2 \prod_{{\bar k}^\prime\ne{\bar k}} (z^2 - \Omega_{{\bar k}^\prime}^2) = \prod_\nu (z^2 - \omega_\nu^2).
\ee
[Here $\bar\Omega_0^2\equiv\Omega_0^2 + \sum_k \alpha_k^2/ \Omega_k^2$ is the square of the ``bare" system frequency in Eq.(3).]  Set $z\rightarrow \Omega_k$ and use the product representation Eq.(6) for $g(z)$ to see that the term in question is unity.

\medskip
I thank Tom\'as Arias, Piet Brouwer, Jan von Delft, Jim Sethna, and Ben Widom for useful comments.  Selman Hershfield made helpful suggestions; he and his colleagues arranged a most pleasant four week visit to the University of Florida, where this work was in part done.


\begin{thebibliography}{99}

\bibitem{BK}
L. P. Kadanoff and G Baym, ``Quantum statistical mechanics; Green's function methods in equilibrium and nonequilibrium problems," W. A. Benjamin (New York) 1962.

\bibitem{LK}
L. V. Keldysh, Zh. Eksp. Teor. Fiz. {\bf 47}, 1515 (1964) [Sov. Phys.---JETP {\bf 20}, 1018 (1965)].

\bibitem{RS}
 J. Rammer and H. Smith, Rev. Mod. Phys. {\bf 58}, 323 (1986).
 
 \bibitem{DL}
D. C. Langreth, 1975 NATO Advanced Study Institute on Linear and 
Nonlinear Electron Transport in Solids,  J. T. Devreese and V. E. van Doren (eds.), p. 3, Plenum (New York) 1976.

\bibitem{AK}
A. Kamenev, in  ``Nanophysics: Coherence and Transport,''
H. Bouchiat, et al. (eds.), p. 177, Elsevier (Amsterdam) 2005.

\bibitem{SH}
See, for example, S. Hershfield, Phys. Rev. Lett. {\bf 70}, 2134 (1993); P. W. Brouwer, A. Lamacraft, and K. Flensberg, Phys. Rev. Lett.  {\bf 94}, 136801 (2005).

 \bibitem{LC}
 
An exception is A. O. Caldeira and A. J. Leggett, Physica {\bf 121A}, 587 (1983),where in Eq.(6.35) a result equivalent to a special case---diagonal part of the density matrix; linear dissipation---of the present work is given, but details are omitted on grounds that  the ``procedure is straightforward but extremely tedious."
 
 \bibitem{NGK}
 N. G. van Kampen, J. Stat. Phys. {\bf 115}, 1057 (2004). 
 
 \bibitem{RJR}
 For an early use of this strategy, see R. J. Rubin, J. Am. Chem. Soc. {\bf 90}, 3061 (1968), where many yet earlier references are given.

 
 \bibitem{FLO} 
 G. W. Ford, J. T. Lewis, and R. F O'Connell, J. Stat. Phys. {\bf 53}, 439 (1988). 
 
 \bibitem{VA}
 V. Ambegaokar, J. Stat. Phys., in press; available on-line from www.springerlink.com.
 
 \bibitem{DET}
 
 For a discussion of corrections to the continuum approximation in a particular case, see ref.\cite{RJR} and citations therein.
 
 \end{thebibliography}
\enddocument